\documentclass[english,aps,manuscript,aps,manuscript]{revtex4}
\usepackage[T1]{fontenc}
\usepackage[latin9]{inputenc}
\usepackage{float}
\usepackage{amstext}
\usepackage{amssymb}
\usepackage{graphicx}

\makeatletter
\@ifundefined{textcolor}{}
{%
 \definecolor{BLACK}{gray}{0}
 \definecolor{WHITE}{gray}{1}
 \definecolor{RED}{rgb}{1,0,0}
 \definecolor{GREEN}{rgb}{0,1,0}
 \definecolor{BLUE}{rgb}{0,0,1}
 \definecolor{CYAN}{cmyk}{1,0,0,0}
 \definecolor{MAGENTA}{cmyk}{0,1,0,0}
 \definecolor{YELLOW}{cmyk}{0,0,1,0}
}

\makeatother

\usepackage{babel}
\begin{document}

\title{Decoherence reduction via continuous dynamical decoupling: Analytical
study of the role of the noise spectrum}

\author{J.M. Gomez Llorente, I. Gomez-Ojeda, and J. Plata}

\address{Departamento de F\'{\i}sica and IUdEA, Universidad de La Laguna,\\
 La Laguna E38200, Tenerife, Spain.}
\begin{abstract}
We analyze the robust character against non-static noise of \emph{clock
transitions} implemented via a method of continuous dynamical decoupling
(CDD) in a hyperfine Zeeman multiplet in $^{87}\textrm{Rb}$. The
emergence of features specific to the quadratic corrections to the
linear Zeeman effect is evaluated. Our analytical approach, which
combines methods of stochastic analysis with time-dependent perturbation
theory, allows tracing the decoherence process for generic noise sources.
Working first with a basic CDD scheme, it is shown that the amplitude
and frequency of the (sinusoidal driving) field of control can be
appropriately chosen to force the non-static random input to have
a (time-dependent) perturbative character. Moreover, in the dressed-state
picture, the effect of noise is described in terms of an operative
random variable whose properties, dependent on the driving field,
can be analytically characterized. In this framework, the relevance
of the spectral density of the fluctuations to the performance of
the CDD technique is precisely assessed. In particular, the range
of noise correlation times where the method of decoherence reduction
is still efficient is identified. The results obtained in the basic
CDD framework are extrapolated to concatenated schemes. The generality
of our approach allows its applicability beyond the specific atomic
system considered.
\end{abstract}
\maketitle

\section{Introduction}

Decoherence in a quantum system, i.e., the loss of \emph{purity} generated
by the coupling to non-controllable environments, is a fundamental
difficulty in the realization of intrinsically quantum effects. Curbing
the effect of the interactions of the system components with the environment,
and, consequently, extending the coherence times is a basic requirement
for the advances in the implementation of quantum technologies \cite{key-Preskill,key-Balasubramanian,key-Lukin,key-Plenio3,key-Metrology}.
Indeed, a primary objective of the research in this field is the development
of technical schemes\textbf{ }to steer the system evolution while
protecting the relative phases. Apart from technical importance, preserving
the coherence has central relevance to fundamental areas of research.
In this sense, it is worth pointing out its crucial role in the realization
of fundamental effects with ultracold atoms \cite{key-Spielman2}. 

Different methods for decoherence reduction have been proposed and
applied in the last decades. Actually, a variety of strategies have
been designed to cope with the specific characteristics of the different
sources of noise. Significant objectives have been achieved: in some
cases, the coherence times have been enlarged by orders of magnitude.
Among the methods applied, the techniques of dynamical decoupling
stand out as particularly effective. They basically consist in strategies
to effectively disconnect the system from the environment that generates
the fluctuations. Their original design incorporated sequences of
pulses of control devised to average out the effect of noise \cite{key-Hahn,key-Viola,key-Viola2,key-Uhrig}.
In subsequent variations of the original proposals \cite{key-Fanchini,key-Plenio2},
the pulses were replaced by continuous-wave driving fields, aimed
at facilitating the integration of the information protocols and at
simplifying the experimental realization. For those methods to be
operative it is necessary to minimize their (unavoidable) invasive
effect on the system whose control is intended. In this sense, concatenation
schemes set up to deal with the extra noise introduced by the auxiliary
fields have been developed \cite{key-Lidar,key-Plenio,key-BarGill}.
The applicability of those techniques to qubits realized with trapped
ions and atoms, nitrogen vacancies ($\textrm{NV}$) centers in diamond,
or quantum dots has been extensively reported (see for instance \cite{key-Bermudez,key-Monroe,key-Awschalom,key-Paladino}
and references therein). 

Here, we will focus on a recent application of a CDD scheme to atoms
of $\textrm{\textrm{R\ensuremath{\textrm{b}^{87}}}}$ which resulted
in a significant reduction of the effect of magnetic noise on transitions
associated to a hyperfine Zeeman multiplet \cite{key-Spielman}. Indeed,
a system of \emph{clock transitions} almost immune to the presence
of noise was generated. The attained stability against fluctuations
has played a key role in subsequent research on the implementation
of different fundamental effects with Bose-Einstein condensates of
$\textrm{\textrm{R\ensuremath{\textrm{b}^{87}}}}$ \cite{key-Spielman3}.
The technique applied to build up the CDD scheme was based on using
a radio-frequency driving field orthogonal to the original Zeeman
component. The objective was to force the magnetic-field fluctuations
to play a secondary role in the dynamics. To cope with the additional
noise introduced via stochastic variations in the driving intensity,
a concatenation scheme was incorporated: a second field of control
was designed to mitigate the effect of that extra noisy input. In
the analysis of the experimental realization, the noise was assumed
to be static: no time dependence of the fluctuations was contemplated.
Hence, the random changes in the magnetic field were considered to
merely affect the reproducibility of the initial conditions for the
different experimental realizations. Here, it is worth stressing that
it is in slow-noise setups where the performance of the CDD techniques
have been mainly evaluated. Moreover, in the studies where non-static
noise has been contemplated, its effect has been frequently analyzed
via numerical simulation or through approximations valid only in specific
regimes (the limit of large observation times or the adiabatic scenario
have been usually tackled). It is pertinent to add that the majority
of those studies have dealt with the pulsed variant of the dynamical-decoupling
technique \cite{key-Alvarez}. In our work, we will go beyond that
scenario: the potential applicability of the CDD method to deal with
generic fluctuations will be analytically evaluated. In particular,
the robustness against non-static noise of the \emph{clock transitions}
implemented in \cite{key-Spielman} will be assessed. Our approach
starts with a description of the system dynamics previous to the application
of the CDD method: the dephasing effects of generic magnetic fluctuations
on the Zeeman multiplet will be analytically described, the emphasis
being put on the differential features associated to the spectrum
and correlation time of noise. Then, to analyze how the system dynamics
is modified in the CDD scheme, we will work with the basis of eigenstates
of the driving term (the so called \emph{dressed-state} representation).
It will be apparent that the magnetic-noise component, diagonal in
the original Zeeman-state basis, becomes off-diagonal in the new representation.
Moreover, for a sufficiently large driving intensity, and, consequently,
for a large dressed-energy separation, the (non static) noise contribution
to the dynamics can be regarded as a (time-dependent) perturbative
term. The use of methods of stochastic analysis combined with the
application of time-dependent perturbation theory will allow us to
characterize the efficiency of the CDD method. In particular, the
relevance of the noise spectral density to the performance of the
decoherence-reduction technique will be studied. Some practical conclusions
on extending the range of applicability of the method will be extracted
from our results.

The outline of the paper is as follows. In Sec. II, we analyze some
fundamental aspects of the decohering effects of generic noise on
a hyperfine Zeeman multiplet. An approach of complete validity will
allow us to trace general dephasing features emergent in the asymptotic
regimes. Additionally, we will describe the loss of \emph{purity}
in any time regime for fluctuations potentially relevant to different
experimental setups. In Sec. III, the system dynamics in the CDD scheme
will be tackled. The analytical characterization of the noise-induced
transfer of population between dressed states will be used to scrutinize
the stability of the \emph{clock transitions} implemented in \cite{key-Spielman}.
First, we will concentrate on the linear Zeeman regime associated
to weak magnetic fields. Then, in Sec. IV, it will be shown that quadratic
corrections to the linear Zeeman effect do not alter the operative
character of the CDD technique. As a proof of consistency, we will
recover the findings of previous work on slow-noise by taking the
limit of large correlation time in our results. The connection with
former predictions on the role of the noise spectrum in the CDD-method
performance will be also established. Finally, the general conclusions
are summarized in Sec. V.

\section{The effect of magnetic fluctuations on the coherent evolution in
a Zeeman multiplet: The relevance of specific noise properties}

As in Ref. \cite{key-Spielman}, we consider here the system formed
by the three $m_{F}$ states of the $F=1$ ground-state manifold of
$\textrm{R\ensuremath{\textrm{b}^{87}}}$ (electronic configuration
$\textrm{[Kr]5\ensuremath{S_{1/2}}}$ and nuclear spin $I=3/2$).
The practical interest of this system is clear: the Zeeman multiplet
resulting from the application of a static magnetic field is a basic
component of arrangements used in a variety of lines of research on
Bose-Einstein condensates. The associated achievements are numerous,
from synthetic spin-orbit coupling \cite{key-Spielman2} to the emulation
of non-Abelian gauge fields \cite{key-Spielman3}, or the generation
of nonlinear Landau-Zener transitions \cite{key-Greene,key-GomezLlorente}.
The applicability of the system demands dealing with the deleterious
effect of magnetic-field noise on the control of the dynamics. Actually,
the fluctuations lead to the broadening of the spectral lines of the
inter-state transitions and induce decoherence in the system evolution.
The study of the noisy dynamics of that system has also general implications
as it exemplifies how the protocols proposed in quantum technologies
can become inefficient due to dephasing in the evolution of the system
components. 

In \cite{key-Spielman}, quadratic corrections to the linear Zeeman
effect were considered. Here, as we focus on the analytical description
of the role of generic noise in the dynamics, we will initially concentrate
on the linear scenario: the differential effects of the noise characteristics
can be already traced in that basic version of the model. Further
on, in Sec. IV, the implications of the quadratic Zeeman effect will
be analyzed. Hence,  we first deal with the Hamiltonian

\begin{equation}
H=\left[\omega_{0}+\delta\omega_{0}(t)\right]F_{z}
\end{equation}
where $\omega_{0}$ denotes the mean value of the characteristic frequency
of the multiplet, $\delta\omega_{0}(t)$ is the shift induced by the
fluctuations, and $\vec{F}$ is the angular momentum operator. In
terms of the system parameters, the mean frequency is expressed as
$\omega_{0}=g_{F}(g_{s}\left|\mu_{B}\right|-g_{I}\mu_{N})B_{0}/\hbar$,
where $g_{F}$ is the Landé factor of the multiplet, $\mu_{B}$ is
the Bohr magneton, $\mu_{N}$ is the nuclear magneton, $g_{s}$ and
$g_{N}$ are, respectively, the $g-$factors of the spin and nuclear
gyro-magnetic ratios. The applied magnetic field $\vec{B}$ will be
expressed as $\vec{B}=[B_{0}+\delta B_{0}(t)]\vec{k}$: $B_{0}$ denotes
the mean value and $\delta B_{0}(t)$ stands for the fluctuations.
The noisy displacement in the frequency,$\delta\omega_{0}(t)$, is
given by $\delta\omega_{0}(t)=g_{F}(g_{s}\left|\mu_{B}\right|-g_{I}\mu_{N})\delta B_{0}(t)/\hbar$.
As any deterministic shift can be included in $\omega_{0},$ $\delta\omega_{0}(t)$
will be considered to have a zero mean value. In the theory developed
to account for the experimental results of \cite{key-Spielman}, the
fluctuations were considered to be static. Hence, it was assumed that
the different realizations of the (time-independent) stochastic variable
$\delta B_{0}$ simply lead to a variation of the initial conditions
for each experimental run. In the present work, we tackle the case
of generic noise; no restrictions on the magnitude of the correlation
time of $\delta\omega_{0}(t)$ are assumed. Our general description
incorporates the static-noise setting as a particular case.

\subsection{The decay of the coherences: general characteristics in the asymptotic
regimes}

In the study of the systems proposed to implement quantum-information
protocols, a quantum description of the environments where decoherence
originates is frequently necessary. In those cases, the theoretical
framework incorporates standard techniques developed in the study
of open quantum systems. In the present case, as noise enters the
system via a classical field, it is feasible to consider the stochastic
variable as a driving element in the system evolution. Accordingly,
the approach used to characterize the dynamics includes as a first
step the analysis of the (unitary) evolution for each \emph{noisy
trajectory}, i.e., for each set of values realized by the random variable
along a time sequence. Subsequently, the statistical average over
noise realizations is carried out \cite{key-Brouard}. 

Our procedure starts by applying the unitary transformation 

\begin{equation}
U(t)=e^{-i\omega_{0}tF_{z}/\hbar}.
\end{equation}
In the associated rotating frame, the system, prepared in the state
$\left|\psi(0)\right\rangle $, evolves, for each stochastic trajectory,
as 

\begin{equation}
\left|\psi(t)\right\rangle =e^{-i\zeta(t)F_{z}/\hbar}\left|\psi(0)\right\rangle ,
\end{equation}
where $\zeta(t)$ is the non-stationary random variable defined by

\begin{equation}
\zeta(t)=\int_{0}^{t}\delta\omega_{0}(t^{\prime})dt^{\prime}.\label{eq:FirstNonstatStochVariable}
\end{equation}
 Correspondingly, the density matrix in the representation of states
$\left\{ \left|k;F,m_{F}\right\rangle \right\} $ ($k$ stands for
additional quantum numbers characterizing the ground-state configuration)
is given by 

\begin{equation}
\rho_{m_{F},m_{F}^{\prime}}(t)=\rho_{m_{F},m_{F}^{\prime}}(0)e^{i(m_{F}-m_{F}^{\prime})\zeta(t)}.\label{eq:}
\end{equation}
 Now, the stochastic character of the system is incorporated by making
the average over fluctuations. The resulting (reduced) density matrix
reads 

\begin{equation}
\rho_{m_{F},m_{F}^{\prime}}(t)=\rho_{m_{F},m_{F}^{\prime}}(0)\left\langle e^{i(m_{F}-m_{F}^{\prime})\zeta(t)}\right\rangle ,\label{eq:expression of the coherences}
\end{equation}
where $\left\langle \right\rangle $ stands for the average over noise
realizations (no confusion with the standard quantum average will
be possible throughout the text). 

From the above equation, it is apparent that the populations do not
change. It is also evident that, to obtain the precise evolution of
the coherences, the statistical characterization of $\zeta(t)$ is
necessary. At this point, a first general difficulty is noticeable:
since $\zeta(t)$ is the sum of elementary increments $\delta\omega_{0}(t)dt$,
which, for finite correlation time $\tau_{c}$, are statistically
dependent, its characterization, and, in turn, the description of
the coherence evolution are not trivial for a generic stochastic variable
$\delta\omega_{0}(t)$. Despite this fundamental limitation, it is
possible to identify important properties of the dephasing, valid
for a generic random input $\delta\omega_{0}(t)$, in the following
regimes:

i) In the limit of large correlation times, i.e., for $t\ll\tau_{c}$,
which corresponds to the slow-noise scenario of the \emph{clock transitions}
implemented in \cite{key-Spielman}, the phase shift can be approximated
as 
\[
\zeta(t)\simeq\delta\omega_{0}(0)t.
\]
 Consequently, the average in Eq. (\ref{eq:expression of the coherences})
is completely determined by the probability distribution $W_{D}[\delta\omega_{0}(0)]$,
i.e., 
\begin{equation}
\rho_{m_{F},m_{F}^{\prime}}(t)=\rho_{m_{F},m_{F}^{\prime}}(0)\int d(\delta\omega_{0})W_{D}(\delta\omega_{0})e^{i(m_{F}-m_{F}^{\prime})\delta\omega_{0}t}.
\end{equation}
In particular, for a Gaussian input $\delta\omega_{0}(0)$ with variance
$\textrm{var}[\delta\omega_{0}]$ we obtain

\begin{equation}
\rho_{m_{F},m_{F}^{\prime}}(t)\propto e^{-\frac{1}{2}(m_{F}-m_{F}^{\prime})^{2}\textrm{var}[\delta\omega_{0}]t^{2}},
\end{equation}
which corresponds to Gaussian decay with characteristic time

\[
\tau_{d}=\left[(m_{F}-m_{F}^{\prime})^{2}\textrm{var(}[\delta\omega_{0})/2\right]^{-1/2}.
\]

ii) In the limit of short correlation times, i.e., for $t\gg\tau_{c}$,
which is eventually reached as longer evolution times are attained
in the monitoring of the system, it is possible to write $\zeta(t)$
in the form

\[
\zeta(t)=\int_{0}^{\Delta t}\delta\omega_{0}(t)dt+\int_{\Delta t}^{2\Delta t}\delta\omega_{0}(t)dt+\cdots+\int_{(n-1)\Delta t}^{t}\delta\omega_{0}(t)dt
\]
with a large $n$, and, still, with an interval $\Delta t$ larger
than $\tau_{c}$, which guarantees that the different summands are
uncorrelated. Hence, applying the Central Limit Theorem \cite{key-Stratonovich},
one concludes that, since $\zeta(t)$ can be expressed as the sum
of a large number of statistically independent variables, it presents
an approximate normal distribution. Therefore one simply needs to
evaluate the mean $\left\langle \zeta(t)\right\rangle $ and the variance
$\left\langle \zeta^{2}(t)\right\rangle -\left\langle \zeta(t)\right\rangle ^{2}$.
Accordingly, we proceed as 

\begin{equation}
\left\langle \zeta(t)\right\rangle =\left\langle \int_{0}^{t}\delta\omega_{0}(t^{\prime})dt^{\prime}\right\rangle =\int_{0}^{t}\left\langle \delta\omega_{0}(t^{\prime})\right\rangle dt^{\prime}=\left\langle \delta\omega_{0}\right\rangle t=0.\label{eq:ShortMean}
\end{equation}
 where we have used the notation $\left\langle \delta\omega_{0}(t)\right\rangle \equiv\left\langle \delta\omega_{0}\right\rangle $
since a stationary input $\delta\omega_{0}(t)$ is being considered.
Now, aiming at the practical applicability of the analysis, we will
evaluate $\left\langle \zeta^{2}(t)\right\rangle $ in terms of a
magnitude of operative use in the characterization of noise, namely,
the spectral density. To this end, we first recall the Wiener-Khinchin
theorem \cite{key-Gardiner}, which connects the Fourier transform
of the autocorrelation function $G(\tau)=\left\langle \delta\omega_{0}(0)\delta\omega_{0}(\tau)\right\rangle $
with the spectrum $S(\omega)$, namely, 

\begin{equation}
S(\omega)=\frac{1}{2\pi}\int_{-\infty}^{\infty}d\tau e^{-i\omega\tau}G(\tau),\label{eq:Spectrum}
\end{equation}
 and the associated inverse expression

\begin{equation}
G(\tau)=\int_{-\infty}^{\infty}d\omega e^{i\omega\tau}S(\omega).\label{eq:AutoCorrelInverse Spectrum}
\end{equation}
 Hence, we use Eq. (\ref{eq:Spectrum}) to calculate the variance
as 

\begin{eqnarray}
\left\langle \zeta^{2}(t)\right\rangle  & = & \left\langle \int_{0}^{t}\delta\omega_{0}(\tau)d\tau\times\int_{0}^{t}\delta\omega_{0}(\tau^{\prime})d\tau^{\prime}\right\rangle \nonumber \\
 & = & \int_{0}^{t}d\tau\int_{0}^{t}d\tau^{\prime}\left\langle \delta\omega_{0}(\tau)\delta\omega_{0}(\tau^{\prime})\right\rangle \nonumber \\
 & = & \int_{-t}^{t}d\tau(t-\left|\tau\right|)\int_{-\infty}^{\infty}d\omega e^{i\omega\tau}S(\omega)\nonumber \\
 & = & 2\int_{0}^{\infty}d\omega\left[\frac{\sin(\omega t/2)}{\omega/2}\right]^{2}S(\omega).\label{eq:VarianceSpectrum}
\end{eqnarray}
 (An appropriate change of variables has been implemented). Furthermore,
in the considered limit $t\gg\tau_{c}$, the function $\left[\frac{\sin(\omega t/2)}{\omega/2}\right]^{2}$
can be approximated in terms of the Dirac delta function $\delta(\omega)$
($\left[\frac{\sin(\omega t/2)}{\omega/2}\right]^{2}\rightarrow2\pi t\delta(\omega)$),
and the integral can be analytically evaluated. Specifically,

\begin{equation}
\left\langle \zeta^{2}(t)\right\rangle \simeq2\pi S(0)t.\label{eq:VarianceZeroSpectrum}
\end{equation}
 It is then concluded that, in the regime considered, the coherences
present an exponential decay, namely, 
\begin{equation}
\rho_{m_{F},m_{F}^{\prime}}(t)\propto e^{-(m_{F}-m_{F}^{\prime})^{2}\pi S(0)t},
\end{equation}
 the $1/e$ scaling time being 
\[
\tau_{d}=\left[(m_{F}-m_{F}^{\prime})^{2}\pi S(0)\right]^{-1}.
\]
 Hence, it is the noise spectrum at zero frequency that determines
the magnitude of the dephasing time. The emergence, irrespective of
the noise properties, of a universal exponential-decay regime in the
limit of long observation times has been analyzed in former studies
on dephasing in different physical contexts \cite{key-Itano,key-Paladino}.
In particular, the dependence of the decay rate on the zero-frequency
spectrum was reported in systems where $\textrm{1/f}$ -noise is relevant.
It is evident that in order to identify the type of noise present
in a particular setup, the results extracted from the analysis of
the asymptotic regimes are not sufficient. Advances in tracking the
fluctuations demand a more complete description of the coherence decay.
A detailed modeling of the noise characteristics is needed to establish
the origin of features emergent in the decoherence process. In the
following, we will proceed along this line.

\subsection{Tracing the dephasing process in a generic time regime}

To describe the system evolution in any time regime, the complete
statistical characterization of $\zeta(t)$ is required, and, consequently,
the properties of $\delta\omega_{0}(t)$ must be specified. Here,
to have a good predictive power in different contexts, a quite general
model with wide practical applicability is assumed. Namely, we consider
that $\delta\omega_{0}(t)$ corresponds to a zero-mean stationary
Ornstein-Uhlenbeck process \cite{key-Gardiner}, i.e., it is a Gaussian
variable whose mean value and correlation function are respectively
given by

\begin{equation}
\left\langle \delta\omega_{0}(t)\right\rangle =0,
\end{equation}
and

\begin{equation}
G(t-t^{\prime})=\left\langle \delta\omega_{0}(t)\delta\omega_{0}(t^{\prime})\right\rangle =\textrm{var}[\delta\omega_{0}]e^{-\alpha\left|t-t^{\prime}\right|},\label{eq:OUautoCorrelation}
\end{equation}
 where $\alpha$ is a positive real coefficient which represents the
inverse of the correlation time, i.e., $\tau_{c}=\alpha^{-1}$. From
Eq. (\ref{eq:Spectrum}), the spectrum is found to be given by 

\begin{equation}
S(\omega)=\frac{\alpha\textrm{var}[\delta\omega_{0}]}{\pi(\alpha^{2}+\omega^{2})}.\label{eq:SpectralDensity}
\end{equation}
 This modeling of noise has been used in previous studies on related
systems \cite{key-Plenio,key-BarGill,key-Brouard2}. In particular,
it was employed in a numerical simulation of the effect of noise on
\emph{clock states} implemented in $\textrm{NV}$ centers in diamond
\cite{key-Stark}. 

The characterization of $\zeta(t)$ follows from the application of
techniques of stochastic analysis \cite{key-Stratonovich}. Specifically,
for the mean value, one has 
\begin{equation}
\left\langle \zeta(t)\right\rangle =\left\langle \int_{0}^{t}\delta\omega_{0}(t^{\prime})dt^{\prime}\right\rangle =\int_{0}^{t}\left\langle \delta\omega_{0}(t^{\prime})\right\rangle dt^{\prime}=0.\label{eq:OU1}
\end{equation}
Additionally, $\left\langle \zeta^{2}(t)\right\rangle $ is obtained
as 

\begin{eqnarray}
\left\langle \zeta^{2}(t)\right\rangle  & = & \left\langle \int_{0}^{t}\delta\omega_{0}(\tau)d\tau\times\int_{0}^{t}\delta\omega_{0}(\tau^{\prime})d\tau^{\prime}\right\rangle \nonumber \\
 & = & \textrm{var}[\delta\omega_{0}]\int_{-t}^{t}d\tau(t-\left|\tau\right|)e^{-\alpha\left|\tau\right|}\nonumber \\
 & = & \frac{2\textrm{var}[\delta\omega_{0}]}{\alpha^{2}}\left(\alpha t+e^{-\alpha t}-1\right)\label{eq:OU2}\\
 & = & 2\pi S(0)\left(t+\frac{e^{-\alpha t}-1}{\alpha}\right)
\end{eqnarray}
 Notice that by fixing $\alpha$ and taking the limits $t\rightarrow0$
and $t\rightarrow\infty$ in this expression, we consistently recover
the results previously obtained using general arguments in the limits
of large correlation time ($t\ll\alpha^{-1}$) and small correlation
time ($t\gg\alpha^{-1}$). In particular, it is shown that, at large
times, $\left\langle \zeta^{2}(t)\right\rangle $ is correctly expressed
as a function of the zero-frequency value of the spectrum $S(0)$.
In the crossover, a complex time dependence, determined by the specific
value of the correlation time, is observed.

Now, once $\left\langle \zeta(t)\right\rangle $ and $\left\langle \zeta^{2}(t)\right\rangle $
are known, the (Gaussian) probability distribution $W_{D}[\zeta(t)]$
is completely determined, and the evolution of the reduced density
matrix is evaluated to give 
\begin{eqnarray}
\rho_{m_{F},m_{F}^{\prime}}(t) & = & \rho_{m_{F},m_{F}^{\prime}}(0)\int d[\zeta(t)]W_{D}[\zeta(t)]e^{i(m_{F}-m_{F}^{\prime})\zeta(t)}\nonumber \\
 & = & \rho_{m_{F},m_{F}^{\prime}}(0)e^{-\frac{1}{2}(m_{F}-m_{F}^{\prime})^{2}\left\langle \zeta^{2}(t)\right\rangle }
\end{eqnarray}

A general remark on the whole system evolution is pertinent. The decay
of the coherences, observed in any time regime and traced when the
statistical average is carried out, reflects the loss of \emph{purity}
in the system evolution. If no coherence-preservation strategies are
implemented, the system is of no use to realize protocols where specific
quantum characteristics are required in large time intervals. We stress
also that the present context corresponds to a phase-fluctuation scenario:
since noise only affects the energy splittings of the used diagonal
representation, it has a purely dephasing effect. There is no loss
of population. This is in contrast with setups where the entrance
of noise occurs through non-diagonal terms (i.e., via terms which
do not commute with the Hamiltonian). There, one speaks of \emph{relaxation}
of the system, instead of pure dephasing. As we will see in the next
section, the present dephasing setting is converted into a relaxation
scenario when the driving field of the CDD is connected.

\section{Application of dynamical-decoupling methods to non-static fluctuations}

The basis of the implementation of the CDD method of \cite{key-Spielman}
was the inclusion in the experimental setup of a driving field orthogonal
to the (static) Zeeman component. To deal with that extra term, a
dressed-state representation, which incorporates the time dependence
of the driving, was used. (See \cite{key-Anderson} for an alternative
scheme which incorporates a continuous-observation scheme). In that
scenario, the eigenvalues and eigenstates of the complete Hamiltonian
can be exactly obtained as the fluctuations are time independent.
Here, in order to build up a framework where the effect of non-static
noise can be tackled, we will resort to a perturbative picture. In
passing, our approach will allow us to clearly identify the basic
mechanism responsible for the effectiveness of the CDD method, and,
in particular, for its functioning in the realization of \cite{key-Spielman}.
In this sense, we point out that, in the dressed-state picture, the
term in the Hamiltonian corresponding to magnetic noise (fluctuations
in $B_{0}$) becomes off-diagonal. Furthermore, for a sufficiently
large separation of the diagonal elements in the new basis, which
can be implemented by increasing the driving intensity, a perturbative
scheme with characteristic parameter given by the quotient between
the noise magnitude and the driving intensity can be set up. Notice
that this procedure is applicable irrespective of the time properties
of the fluctuations. In the case of static noise, the random component
leads to a second-order correction to the eigenvalues. Hence, the
frequencies of the dressed-state transitions become noise immune to
first order. There is a shortcoming in the practical arrangement:
since stochastic variations in the driving intensity cannot be avoided,
the scheme introduces additional fluctuations in the system. Furthermore,
since that extra noisy term enters the diagonal elements, it is a
first-order component of the dressed-state picture. To cope with this
additional random input, a second driving field (the probe field)
orthogonal to the first one is incorporated. The procedure can be
continued: additional concatenated probe fields can be included till
the magnitude of the remnant noise, entering the system through the
last driving field, can be considered to be negligible compared with
the final splitting. 

Let us address now the case of non-static fluctuations. We will concentrate
on a model system that incorporates the basic components of the CDD
schemes. Namely, we will consider the Hamiltonian given by

\begin{equation}
H=\left[\omega_{0}+\delta\omega_{0}(t)\right]F_{z}+2\Omega_{d}\cos(\omega_{d}t)F_{x},\label{eq:HamiltonianCDD}
\end{equation}
where $\Omega_{d}$ and $\omega_{d}$ are the characteristic parameters
of the control field ($\Omega_{d}$ is proportional to the Landé factor
of the hyperfine multiplet and to the field intensity). Only one driving
term is considered: the results obtained for this primary scenario
are straightforwardly generalized to more elaborate arrangements.
(See \cite{key-Garraway} for a proposal of bichromatic dressing).

\subsection{Setting up the perturbative scheme}

Because of the non-static character of $\delta\omega_{0}(t)$, the
method used in previous work \cite{key-Spielman} to analytically
characterize the system dynamics is not applicable: it is not possible
to obtain exact eigenvalues of $H$. Still, an alternative procedure
to evaluate the performance of the CDD scheme can be devised. Namely,
by choosing the driving frequency as $\omega_{d}=\omega_{0}$, and
working in the rotating frame defined by the unitary transformation

\begin{equation}
U_{1}(t)=e^{-i\omega_{d}tF_{z}/\hbar},\label{eq:U1}
\end{equation}
the Hamiltonian in Eq. (\ref{eq:HamiltonianCDD}) is rewritten as

\begin{equation}
H=\delta\omega_{0}(t)F_{z}+\Omega_{d}F_{x},
\end{equation}
where the Rotating Wave Approximation (RWA) has been applied ($\Omega_{d}$
is assumed to be much smaller than $\omega_{d}$) and the same notation
$H$ is being used for the rotated Hamiltonian $U_{1}^{\dagger}HU_{1}-i\hbar U_{1}^{\dagger}\dot{U}_{1}$.
The additional transformation

\begin{equation}
U_{2}(t)=e^{-i\frac{\pi}{2}F_{y}/\hbar}\label{eq:U2}
\end{equation}
leads to

\begin{equation}
H=\Omega_{d}F_{z}+\delta\omega_{0}(t)F_{x}.\label{eq:RotatedHamiltCDD}
\end{equation}

From the form of $H,$ it is apparent that, if the driving-field intensity
is much stronger than the noise magnitude, i.e., for $\Omega_{d}\gg\left|\delta\omega_{0}(t)\right|$,
the Hamiltonian can be split as the sum of a zero-order term 
\[
H_{0}=\Omega_{d}F_{z}
\]
(with eigenstates $\left|F,\tilde{m}_{F}\right\rangle $, $\tilde{m}_{F}=0,\pm1$,
and associated eigenvalues $E(\tilde{m}_{F})=\tilde{m}_{F}\Omega_{d}\hbar$)
and a (time-dependent) perturbative contribution 
\begin{equation}
W(t)=\delta\omega_{0}(t)F_{x}.\label{eq:timedependentPerturbation}
\end{equation}
In this approach, the effect of $W(t)$ for each noisy trajectory
can be characterized. From time-dependent perturbation theory, it
is known that, for the system prepared in one of the states, let us
say the state $\left|\tilde{m}_{F}\right\rangle $, the probability
of transition to other state ($\left|\tilde{m}_{F}^{\prime}\right\rangle $)
is given to first-order by

\begin{eqnarray}
P_{\tilde{m}_{F},\tilde{m}_{F}^{\prime}}(t) & = & \frac{1}{\hbar^{2}}\left|\int_{0}^{t}dt^{\prime}W_{\tilde{m}_{F}^{\prime},\tilde{m}_{F}}(t^{\prime})e^{i(E_{\tilde{m}_{F}^{\prime}}-E_{\tilde{m}_{F}})t^{\prime}/\hbar}\right|^{2}\nonumber \\
 & = & \frac{\left|(F_{x})_{\tilde{m}_{F}^{\prime},\tilde{m}_{F}}\right|^{2}}{\hbar^{2}}\left|\int_{0}^{t}dt^{\prime}\delta\omega_{0}(t^{\prime})e^{i(\tilde{m}_{F}^{\prime}-\tilde{m}_{F})\Omega_{d}t^{\prime}}\right|^{2}
\end{eqnarray}
where it has been taken into account that, since $E_{\tilde{m}_{F}^{\prime}}$
and $E_{\tilde{m}_{F}}$ are zero-order eigenvalues, their difference
is given by $E_{\tilde{m}_{F}^{\prime}}-E_{\tilde{m}_{F}}=(\tilde{m}_{F}^{\prime}-\tilde{m}_{F})\Omega_{d}\hbar$
(see the form of $H_{0}$). Additionally, the expression of the matrix
element 
\[
W_{\tilde{m}_{F}^{\prime},\tilde{m}_{F}}(t^{\prime})=\delta\omega_{0}(t^{\prime})(F_{x})_{\tilde{m}_{F}^{\prime},\tilde{m}_{F}}
\]
has been used. The next step is dealing with the stochastic character
of the evolution. Let us see that the statistical average

\begin{eqnarray}
\left\langle P_{\tilde{m}_{F},\tilde{m}_{F}^{\prime}}(t)\right\rangle  & = & \frac{\left|(F_{x})_{\tilde{m}_{F}^{\prime},\tilde{m}_{F}}\right|^{2}}{\hbar^{2}}\left\langle \left|\int_{0}^{t}dt^{\prime}\delta\omega_{0}(t^{\prime})e^{i(\tilde{m}_{F}^{\prime}-\tilde{m}_{F})\Omega_{d}t^{\prime}}\right|^{2}\right\rangle \label{eq:PopulationTransferNoisy}
\end{eqnarray}
is a useful indicator of the efficiency of the CDD method. From a
first qualitative evaluation, one can conclude that, when the driving
intensity is increased, more rapid does become the oscillation resulting
from the exponential $e^{i(\tilde{m}_{F}^{\prime}-\tilde{m}_{F})\Omega_{d}t^{\prime}}$.
Consequently, provided that $\delta\omega_{0}(t^{\prime})$ has not
harmonic components in resonance with $e^{i(\tilde{m}_{F}^{\prime}-\tilde{m}_{F})\Omega_{d}t^{\prime}}$,
an effective averaging out of the integral value can be predicted.
This consideration can also be formulated from the statistical analysis
of the stochastic variable defined as 

\begin{equation}
\chi(t)=\int_{0}^{t}dt^{\prime}\delta\omega_{0}(t^{\prime})e^{i(\tilde{m}_{F}^{\prime}-\tilde{m}_{F})\Omega_{d}t^{\prime}},
\end{equation}
present in Eq. (\ref{eq:PopulationTransferNoisy}). It is shown that,
if the spectral density of $\delta\omega_{0}(t^{\prime})$ does not
reach a significant value at the frequency $(\tilde{m}_{F}^{\prime}-\tilde{m}_{F})\Omega_{d}$,
the exponential factor $e^{i(\tilde{m}_{F}^{\prime}-\tilde{m}_{F})\Omega_{d}t^{\prime}}$
leads to a reduction in the variance of $\chi(t)$ with respect to
that of the non-modulated variable $\zeta(t)$ given by Eq. (\ref{eq:FirstNonstatStochVariable}).
Consequently, in that case, the inhibition of the population transfer
as $\Omega_{d}$ grows can be conjectured. In the following, we will
see that a quantitative analysis confirms these predictions.

\subsection{Incorporating the noise characteristics}

Taking into account the Ornstein-Uhlenbeck characteristics of $\delta\omega_{0}(t)$,
the average of the population transfer given by Eq. (\ref{eq:PopulationTransferNoisy})
is evaluated as follows 

\begin{eqnarray}
\left\langle P_{\tilde{m}_{F},\tilde{m}_{F}^{\prime}}(t)\right\rangle  & = & \frac{\left|(F_{x})_{\tilde{m}_{F}^{\prime},\tilde{m}_{F}}\right|^{2}}{\hbar^{2}}\left\langle \int_{0}^{t}d\tau\delta\omega_{0}(\tau)e^{i\Omega_{e}\tau}\times\int_{0}^{t}d\tau^{\prime}\delta\omega_{0}(\tau^{\prime})e^{-i\Omega_{e}\tau^{\prime}}\right\rangle \nonumber \\
 & \propto & \textrm{var}[\delta\omega_{0}]\int_{0}^{t}d\tau\int_{0}^{t}d\tau^{\prime}e^{-\alpha\left|\tau-\tau^{\prime}\right|}e^{i\Omega_{e}(\tau-\tau^{\prime})}\nonumber \\
 & = & \frac{2\textrm{var}[\delta\omega_{0}]}{\alpha^{2}+\Omega_{e}^{2}}\left[\alpha t+\frac{\Omega_{e}^{2}-\alpha^{2}}{\alpha^{2}+\Omega_{e}^{2}}(1-e^{-\alpha t}\cos\Omega_{e}t)-\frac{2\alpha\Omega_{e}e^{-\alpha t}}{\alpha^{2}+\Omega_{e}^{2}}\sin\Omega_{e}t\right]\label{eq:FinalPopulTransfer}
\end{eqnarray}
 where we have used the effective frequency $\Omega_{e}=(\tilde{m}_{F}^{\prime}-\tilde{m}_{F})\Omega_{d}$,
and, the integral has been calculated via an adequate change of variables. 

From the above expression some preliminary conclusions can be drawn:

i) A crucial aspect of the applicability of the CDD method is uncovered
by the analysis of the regime $\alpha\gg\Omega_{e}$. It is apparent
from Eq. (\ref{eq:FinalPopulTransfer}) that the role of the field
intensity $\Omega_{d}$ loses relevance as the noise correlation time
decreases, i.e., for a growing $\alpha$. Furthermore, for $\alpha\gg\Omega_{e}$,
the dependence of the population transfer on $\Omega_{d}$ vanishes.
This finding can be understood using arguments relative to the spectral
decomposition of the fluctuations. In this sense, it is convenient
to work with the Fourier transform of the stochastic variable $\delta\omega_{0}(t)$.
Accordingly, we write

\begin{equation}
\delta\omega_{0}(t)=\int d\omega c(\omega)e^{i\omega t}
\end{equation}
 where the harmonic components are given by 

\begin{equation}
c(\omega)=\frac{1}{2\pi}\int dt\delta\omega_{0}(t)e^{-i\omega t},
\end{equation}
 and are distributed according to the (Lorentzian) spectral density
given by Eq. (\ref{eq:SpectralDensity}). {[}See Ref. \cite{key-Gardiner})
for a complete statistical characterization of $c(\omega)${]}. Notice
that, as the correlation time decreases, the spectrum becomes wider. 

Hence, using the harmonic components, the perturbation can be rewritten
as 

\[
W(t)=\left(\int d\omega c(\omega)e^{i\omega t}\right)F_{x},
\]
 and the Hamiltonian in Eq. (\ref{eq:RotatedHamiltCDD}) can be regarded
as representing the driving of the (dressed) triplet system by a pulse
of harmonic signals which are effective in inducing inter-state transitions
only when the quasi-resonance condition $\omega\simeq\Omega_{e}$
is fulfilled. Eq. (\ref{eq:SpectralDensity}) makes it evident that,
for small values of $\alpha$ (large correlation times), i.e., for
a narrow spectrum, there are no harmonic components of noise in resonance
with the inter-state transition frequency $\Omega_{e}$. Hence, the
noise-induced transfer of population is blocked. On the other hand,
for a sufficiently large value of $\alpha$, and, in turn, for a large
spectral width, the variation of $\Omega_{e}$ does not reduce the
fraction of noisy components in resonance with that frequency. In
that stage, the value of the energy splitting is not longer a limiting
element of the population transfer. It is then understood that in
the range defined by $\alpha\gg\Omega_{e}$, the CDD method is not
longer effective for mitigating the effect of the fluctuations. These
findings are illustrated in Figs. 1 and 2, where the transition probability
is represented as a function of time for different sets of parameters
$\alpha$ and $\Omega_{e}$. Notice that the CDD scheme is highly
efficient for small values of $\alpha$ (Fig. 1): the transition probability
is significantly reduced as the driving intensity is increased. In
contrast, the differential effect of the CDD scheme for growing driving
intensity is hardly noticeable for a wide spectral density, i.e.,
for a large $\alpha$ (Fig. 2). (In order to focus on the combined
role of the noise spectral width and the driving intensity, we have
used the scale factor $A=\frac{\left|(F_{x})_{\tilde{m}_{F}^{\prime},\tilde{m}_{F}}\right|^{2}}{\hbar^{2}}\textrm{var}[\delta\omega_{0}]$
in the representation of the transition probability).

ii) Additional arguments in the same line are extracted by expressing
the population transfer, given by Eq. (\ref{eq:FinalPopulTransfer}),
as a function of the spectrum, i.e., 

\begin{eqnarray}
\left\langle P_{\tilde{m}_{F},\tilde{m}_{F}^{\prime}}(t)\right\rangle  & \propto & S(\Omega_{e})\left[t+\frac{1}{\alpha}\frac{\Omega_{e}^{2}-\alpha^{2}}{\alpha^{2}+\Omega_{e}^{2}}(1-e^{-\alpha t}\cos\Omega_{e}t)-\frac{2\Omega_{e}e^{-\alpha t}}{\alpha^{2}+\Omega_{e}^{2}}\sin\Omega_{e}t\right].\label{eq:FinalPopulSpectrum}
\end{eqnarray}
 We stress that no divergence emerges from the term that incorporates
the factor $\frac{1}{\alpha}$: as can be seen in Eq. (\ref{eq:SpectralDensity}),
the spectral density $S(\Omega_{e})$ includes a factor $\alpha$. 

The central role played by the spectral component corresponding to
the effective frequency $S(\Omega_{e})$ is evident in Eq. (\ref{eq:FinalPopulSpectrum}).
For a narrow spectrum, it is possible, by increasing the driving intensity,
to force $\Omega_{e}$ out of the spectral range, i.e., to make $S(\Omega_{e})\simeq0$.
In contrast, for a flat spectrum, no significant variations in $S(\Omega_{e})$
take place as $\Omega_{e}$ grows. 

iii) Although it is already patent in the above arguments, it is worth
stressing that, except in the regime $\alpha\gg\Omega_{e}$, the CDD
method can be considered to be efficient. Indeed, as the driving intensity,
proportional to $\Omega_{d}$, and, consequently to $\Omega_{e}$,
is increased, the effect of the stochastic input is reduced: the averaged
probability of transition diminishes as $\Omega_{e}$ grows, and,
therefore, the zero-order eigenstates (the dressed states $\left|F,\tilde{m}_{F}\right\rangle ,\,\tilde{m}_{F}=0,\pm1$)
approximate better the eigenstates of the complete noisy Hamiltonian.

iv) Another proof of consistency of the whole approach is obtained
by checking that the results corresponding to static fluctuations
are recovered in the limit of large correlation times. For $\tau_{c}\rightarrow\infty\,(\alpha\rightarrow0)$,
the transfer of population is given by 

\begin{equation}
\left\langle P_{\tilde{m}_{F},\tilde{m}_{F}^{\prime}}(t)\right\rangle =\frac{\left|(F_{x})_{\tilde{m}_{F}^{\prime},\tilde{m}_{F}}\right|^{2}}{\hbar^{2}}\frac{2\textrm{var}[\delta\omega_{0}]}{\Omega_{e}^{2}}\left(1-\cos\Omega_{e}t\right),
\end{equation}
 which matches the average over noise realizations of the probability
of transition between two dressed states induced by a static random
perturbation $W=\delta\omega_{0}F_{x}$. Note that the order of magnitude
of this first perturbative correction is determined by the quotient
$\textrm{var}[\delta\omega_{0}]/\Omega_{e}^{2}$, in agreement with
the precision reached in the application of CDD methods to static
noise \cite{key-Spielman}. Therefore, the zero-order eigenstates
(the dressed states of the used representation) are approximate eigenstates
of the complete (noisy) Hamiltonian, with precision given by $\textrm{var}[\delta\omega_{0}]/\Omega_{e}^{2}$.
Using the terminology introduced in \cite{key-Spielman}, one recovers
the conclusion that the \emph{clock states} are noise immune to first
order in the quotient $\delta\omega_{0}/\Omega_{e}$.

v) The above conclusions, extracted from the study of the basic CDD
method, i.e., for the scheme incorporating one driving field, are
straightforwardly extrapolated to more elaborate setups. As, in any
stage in the CDD scheme, the last noisy component entering the system
is transferred to an off-diagonal term through an appropriate change
of representation, its effect on the dynamics can always be characterized
in terms of a population transfer between effective zero-order eigenstates
similar to that given by Eqs. (\ref{eq:PopulationTransferNoisy}).
Therefore, the effectiveness of the decoherence-reduction method is
guaranteed provided that the \emph{final} inter-state transition frequencies
are out of the dominant part of the spectral range of the residual
noise. Note that controlling the frequencies of transition, in particular,
the effective frequency $\Omega_{e}$, to avoid the occurrence of
resonances with the noise spectral components has the limitations
associated to the application of the RWA and to the system reduction
employed in the description of the model system. A careful analysis
of each experimental setup is needed: since the consecutive application
of the RWA as different drivings are incorporated implies a reduction
in the magnitude of the splittings, keeping the last $\Omega_{e}$
outside the spectral range of the corresponding final noise is not
trivial.

vi) It is worth pointing out that, since the application of time-dependent
perturbation theory to first-order requires only up to the second
moment of noise, the used framework embodies in fact a Gaussian approximation. 

\begin{figure}[H]
\centerline{\includegraphics{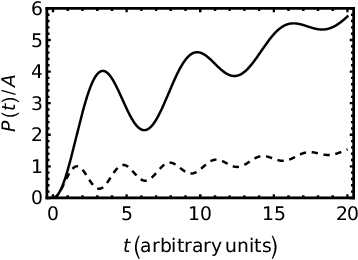}}\caption{Noise-induced transition probability $P$ between two dressed states
as a function of time (in arbitrary units). The used parameters are
$\alpha=0.1$ and $\Omega_{e}=1$ (continuous line), and, $\alpha=0.1$
and $\Omega_{e}=2$ (dashed line). (We have used the scale factor
$A=\frac{\left|(F_{x})_{\tilde{m}_{F}^{\prime},\tilde{m}_{F}}\right|^{2}}{\hbar^{2}}\textrm{var}[\delta\omega_{0}]$
in the representation of the probability).}
\end{figure}

\begin{figure}[H]
\centerline{\includegraphics{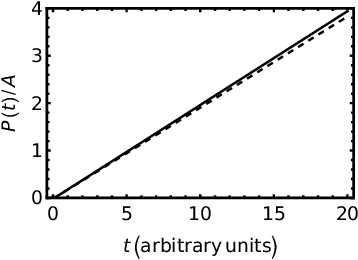}}\caption{Same caption as that of Fig. 1, with $\alpha=10$ and $\Omega_{e}=1$
(continuous line), and, $\alpha=10$ and $\Omega_{e}=2$ (dashed line). }
\end{figure}

\section{The role of the quadratic Zeeman Effect in the dynamical-decoupling
scheme}

Given the high precision required for the characterization of noise
effects, the system description must go beyond the strictly linear
regime corresponding to the Zeeman effect in the weak magnetic-field
limit. The next-order contribution to the Zeeman shift accounts for
the field-induced coupling between hyperfine multiplets. For alkali
gases, that correction can be analytically evaluated (the simultaneous
matrix representation of both, hyperfine and Zeeman, terms can be
decomposed into blocks characterized by the value of the quantum number
$m_{F}$ which can be analytically solved). Indeed, using the Breit-Rabi
formula \cite{key-BreitRabi,key-Stamper-Kurn}, the net quadratic
Zeeman shift can be approximated as 

\begin{equation}
\hbar\epsilon\left(\frac{F_{z}^{2}}{\hbar^{2}}-I\right)
\end{equation}
where the parameter $\epsilon$ is given by 
\[
\epsilon=\frac{(g_{s}\mu_{B}-g_{I}\mu_{N})^{2}B^{2}}{4\Delta W_{hf}},
\]
with $\Delta W_{hf}$ being the hyperfine energy splitting between
the $F=2$ and $F=1$ terms. (We stress that a more accurate approximation
to the quadratic corrections does not alter the conclusions of this
section). Accordingly, the Hamiltonian that describes the original
Zeeman multiplet is rewritten as

\begin{equation}
H=\left[\omega_{0}+\delta\omega_{0}(t)\right]F_{z}+\hbar\epsilon\left(\frac{F_{z}^{2}}{\hbar^{2}}-I\right).
\end{equation}
 Important for the evaluation of the efficiency of the decoupling
scheme, is to keep in mind that the dominant noisy contribution comes
from the linear dependence of $\omega_{0}$ on the random magnetic
field. Given its second-order character, the stochastic variation
of $\epsilon$ will not be included in the model. 

The system evolution for each stochastic trajectory is given by 

\begin{equation}
\left|\psi(t)\right\rangle =e^{-i\left[\zeta(t)F_{z}/\hbar+\epsilon t\left(F_{z}^{2}/\hbar^{2}-I\right)\right]}\left|\psi(0)\right\rangle ,
\end{equation}
 which, in terms of the density matrix, and incorporating the statistical
average, reads 

\begin{equation}
\rho_{m_{F},m_{F}^{\prime}}(t)=\rho_{m_{F},m_{F}^{\prime}}(0)e^{i(m_{F}^{2}-m_{F}^{\prime2}-2)\epsilon t}\left\langle e^{i(m_{F}-m_{F}^{\prime})\zeta(t)}\right\rangle ,\label{eq:expression of the coherencesQUADRATIC}
\end{equation}
 where $\zeta(t)$ is still given by Ec. (\ref{eq:FirstNonstatStochVariable}).
As the considerations made in Sec. II on the statistical average are
still applicable, it is concluded that the quadratic corrections to
the linear Zeeman effect merely leads to an oscillation of the coherences
as their decay proceeds. The possibility of observing that oscillation
in the experiments depends on the relative magnitude of the parameter
$\epsilon$ and the dephasing rate evaluated in Sec. II.

When the CDD scheme is applied and the field of control is connected,
the Hamiltonian that governs the dynamics reads

\begin{equation}
H=\left[\omega_{0}+\delta\omega_{0}(t)\right]F_{z}+\hbar\epsilon\left(\frac{F_{z}^{2}}{\hbar^{2}}-I\right)+2\Omega_{d}\cos(\omega_{d}t)F_{x},
\end{equation}
 which, through the sequential application of the unitary transformations
given by Eqs. (\ref{eq:U1}) and (\ref{eq:U2}), is cast into the
form 

\[
H=\Omega_{d}F_{z}+\hbar\epsilon\left(\frac{F_{x}^{2}}{\hbar^{2}}-I\right)+\delta\omega_{0}(t)F_{x}.
\]
Hence, the previous perturbative scheme must be redefined. Whereas
the zero-order Hamiltonian is now given by 

\[
H_{0}=\Omega_{d}F_{z}+\hbar\epsilon\left(\frac{F_{x}^{2}}{\hbar^{2}}-I\right),
\]
 the perturbation still corresponds to Ec. (\ref{eq:timedependentPerturbation}).
The zero-order eigenvalues $E_{\xi}$, where $\xi=x,y,z$, are straightforwardly
obtained:

\begin{eqnarray}
\frac{E_{x}}{\hbar}=\omega_{x} & = & 0,\nonumber \\
\frac{E_{y}}{\hbar}=\omega_{y} & = & \frac{-\epsilon+\sqrt{\epsilon^{2}+4\Omega_{d}^{2}}}{2},\nonumber \\
\frac{E_{z}}{\hbar}=\omega_{z} & = & -\frac{\epsilon+\sqrt{\epsilon^{2}+4\Omega_{d}^{2}}}{2}.\label{eq:eigenvalues clock states}
\end{eqnarray}
\medskip{}
\medskip{}
\medskip{}
 Moreover, the associated eigenstates $\left|x\right\rangle $, $\left|y\right\rangle $,
and, $\left|z\right\rangle $ (the notation refers to the analogy
existent with the states of the Cartesian basis \cite{key-Spielman})
can be written as

\begin{equation}
\left|\xi\right\rangle =c_{\xi,1}\left|1,1\right\rangle +c_{\xi,0}\left|1,0\right\rangle +c_{\xi,-1}\left|1,-1\right\rangle ,\label{eq:clock eigenstates general expression}
\end{equation}
with

\begin{eqnarray}
c_{\xi,1} & = & \left[2+\frac{4\omega_{\xi}^{2}}{\Omega_{d}^{4}}(\omega_{\xi}+\epsilon)^{2}-\frac{4\omega_{\xi}}{\Omega_{d}^{2}}(\omega_{\xi}/2+\epsilon)\right]^{-1/2}\nonumber \\
c_{\xi,0} & = & \frac{\sqrt{2}\omega_{\xi}}{\Omega_{d}}c_{\xi,1}\label{eq:SecondCoeffEigenstate}\\
c_{\xi,-1} & = & -\left[1-\frac{2\omega_{\xi}}{\Omega_{d}^{2}}(\omega_{\xi}+\epsilon)\right]c_{\xi,1}\label{eq:ThirdCoeffEigenstate-1}
\end{eqnarray}
Now, time-dependent perturbation theory can be directly applied and
the probability of noise-induced transition between two of the dressed
states (let us say $\left|\xi\right\rangle $ and $\left|\xi^{\prime}\right\rangle $)
is given by 

\begin{eqnarray}
\left\langle P_{\xi,\xi^{\prime}}(t)\right\rangle  & = & \frac{\left|(F_{x})_{\xi^{\prime},\xi}\right|^{2}}{\hbar^{2}}\left\langle \left|\int_{0}^{t}dt^{\prime}\delta\omega_{0}(t^{\prime})e^{i(\omega_{\xi^{\prime}}-\omega_{\xi})t^{\prime}}\right|^{2}\right\rangle .\label{eq:PopulationTransferNoisyQuadratic}
\end{eqnarray}

From the analogy of this expression with Eq. (\ref{eq:PopulationTransferNoisy}),
it follows that the considerations of the previous section on the
reduction of the noise effects achieved in the linear-Zeeman regime
by using the CDD method are still applicable. The quadratic corrections
modify the zero-order eigenvalues of the perturbative scheme, and,
consequently, the resonance condition between the transition frequencies
and the noisy harmonic components which determines the effectiveness
of the method. However, they do not affect the functioning of the
dynamical-decoupling mechanism. Note that, as, now, the zero-order
energy levels are not equally spaced, the efficiency of the fluctuations
to induce transitions can be dependent on the specific two states
involved: the values of the noise spectrum at the different transition
frequencies can present a non-negligible variation.

\section{Concluding remarks}

Given the variety of sources of noise that can be relevant to the
experimental setups, a realistic consideration of the applicability
of CDD methods should contemplate the potential role of finite correlation
times. Indeed, it is sensible to go beyond a scenario where all the
fluctuations (the original input and those resulting from random variations
of the different auxiliary fields) are considered to be static. The
present study provides some clues to deal with that issue: we have
rigorously shown that the use of simplified static-noise models is
appropriate as far as the inter-state transition frequencies are outside
the dominant spectral ranges of the fluctuations. Whereas, previous
to the application of the CDD method, it is the zero-frequency value
of the noise spectrum that determines the asymptotic dephasing rate,
in the CDD setup, the decoherence time is basically determined by
the noise spectrum at the final effective frequency $\Omega_{e}$.
Decoherence is significantly reduced if $\Omega_{e}$ does not enter
the relevant part of the spectrum. This is the case of the arrangement
of \cite{key-Spielman}. However, in general, the feasibility of extending
the coherence times by controlling $\Omega_{e}$ is not guaranteed.
Since the application of the RWA at the different stages of the CDD
method implies a reduction in orders of magnitude of $\Omega_{e}$,
reaching a negligible value of the noise spectrum at $\Omega_{e}$
can be problematic as the concatenation scheme proceeds. 

The applicability of the study beyond the considered atomic context
can be envisaged. Indeed, the developed approach is appropriate to
any system that can be effectively described in terms of a zero-order
Hamiltonian and a generic noisy off-diagonal perturbation. Particularly
interesting can be the inclusion of $\textrm{1/f}$-noise in this
framework. We recall that decoherence in qubits implemented with solid-state
devices is frequently studied via a description of the fluctuations
as $\textrm{1/f}$-noise. The compact modeling of the associated correlation
function \cite{key-Stratonovich}, which accounts for the spectrum
form and introduces cutoff frequencies in the spectral range, can
facilitate the application of our approach. Actually, tracing noise-induced
transitions at any time would imply a significant advance in the description
of the decoherence processes in that environment. Another significant
objective is the comparison with the experimental findings to check
the validity of the assumed Gaussian approximation or the mere stationary
character of noise. One can conjecture that the relative magnitude
of the effective transition frequency and the spectral cutoffs must
be central in the performance of the CDD methods with $\textrm{1/f}$
noise.

Finally, it is worth depicting some lines of potential applicability
of the study in noise spectroscopy \cite{key-Hirayama,key-Alvarez2,key-Szankowski,key-Viola4}.
A first general consideration refers to the practical use of the analytical
descriptions of the decoherence processes in the design of proposals
for identifying the fluctuation characteristics. Some specific objectives
can be outlined: the prospect of obtaining the noise spectrum by varying
the effective transition frequency and measuring the asymptotic value
of the decoherence rate seems plausible; indeed, the combination of
that strategy with the information extracted from the analysis of
different time regimes can serve to improve the scrutiny of the noise
properties. The analyticity of the approach allows also the use of
designed random signals to check the precision of the proposals. The
high level of control achieved in the considered experimental setup,
and, in fact, in more general contexts, makes it advisable to employ
the techniques proposed for the realization of quantum information
protocols as elements of noise identification methods.

\section*{Acknowledgments}

One of us (JMGL) acknowledges the support of the Spanish Ministerio
de Economía y Competitividad and the European Regional Development
Fund (Grant No. PID2019-105225GB-I00).

\end{document}